\documentclass[twocolumn,preprintnumbers,amsmath,amssymb]{revtex4}

\usepackage{graphicx}
\usepackage{dcolumn}
\usepackage{bm}
\usepackage{theorem}     
\usepackage{color}        
\usepackage{multirow}    
  
{\theorembodyfont{\normalfont\it}
\theoremheaderfont{\normalfont\bf}
\newtheorem{thm}{ Theorem}
\newtheorem{dfn}[thm]{ Definition}
\newtheorem{lmm}[thm]{ Lemma}

\newtheorem{crl}[thm]{ Corollary}
\newtheorem{asm}[thm]{ Assumption}
\newtheorem{prp}[thm]{ Proposition}
\newtheorem{cjt}[thm]{ Conjecture}
\newtheorem{rmk}[thm]{ Remark}}

{\theorembodyfont{\normalfont}
\theoremheaderfont{\normalfont\it}
\newtheorem{prf}{ Proof:}}

{\theorembodyfont{\normalfont}
\theoremheaderfont{\normalfont\it}
}

{\theorembodyfont{\normalfont}
\theoremheaderfont{\normalfont\it}
}

{\theorembodyfont{\normalfont}
\theoremheaderfont{\normalfont\it}
}

{\theorembodyfont{\normalfont}
\theoremheaderfont{\normalfont\it}
}

{\theorembodyfont{\normalfont}
\theoremheaderfont{\normalfont\it}
}

\newcommand{\bra}[1]{\mbox{$\langle#1|$}}

\newcommand{\ket}[1]{\mbox{$|#1\rangle$}}

\newcommand{\proj}[1]{\mbox{$\ket{#1}\!\bra{#1}$}}

\newcommand{\alg}[1]{\begin{align}#1\end{align}}

\newcommand{\nn}{\nonumber}
\newcommand{\ca}[1]{{\mathcal #1}}
\newcommand{\mbb}[1]{{\mathbb #1}}
\newcommand{\mfk}[1]{{\mathfrak #1}}

\newcommand{\bthm}[1]{\begin{thm}\label{thm:#1}}
\newcommand{\ethm}{\end{thm}}

\newcommand{\rThm}[1]{Theorem \ref{thm:#1}}
\newcommand{\blmm}[1]{\begin{lmm}\label{lmm:#1}}
\newcommand{\elmm}{\end{lmm}}
\newcommand{\rLmm}[1]{Lemma \ref{lmm:#1}}
\newcommand{\rlmm}[1]{\ref{lmm:#1}}
\newcommand{\bdfn}[1]{\begin{dfn}\label{dfn:#1}}
\newcommand{\edfn}{\end{dfn}}

\newcommand{\basm}[1]{\begin{asm}\label{asm:#1}}
\newcommand{\easm}{\end{asm}}

\newcommand{\bprp}[1]{\begin{prp}\label{prp:#1}}
\newcommand{\eprp}{\end{prp}}

\newcommand{\bcrl}[1]{\begin{crl}\label{crl:#1}}
\newcommand{\ecrl}{\end{crl}}

\newcommand{\bcjt}[1]{\begin{cjt}\label{cjt:#1}}
\newcommand{\ecjt}{\end{cjt}}

\newcommand{\brmk}[1]{\begin{rmk}\label{rmk:#1}}
\newcommand{\ermk}{\end{rmk}}

\newcommand{\bprf}{\begin{prf}}
\newcommand{\eprf}{\end{prf}}
\newcommand{\laeq}[1]{\label{eq:#1}}
\newcommand{\req}[1]{(\ref{eq:#1})}

\newcommand{\QED}{\hfill$\blacksquare$}

\newcommand{\lapp}[1]{\label{app:#1}}
\newcommand{\rapp}[1]{\ref{app:#1}}
\newcommand{\rApp}[1]{Appendix \ref{app:#1}}

\newcommand{\bitem}{\begin{itemize}}
\newcommand{\entem}{\end{itemize}}
\newcommand{\benum}{\begin{enumerate}}
\newcommand{\ennum}{\end{enumerate}}

\newcommand{\otm}{\otimes}

\newcommand{\prlsection}[1]{{\it{#1}}.---}

\begin{document}

\title{Gentle Measurement as a Principle of Quantum Theory}

\author{Eyuri Wakakuwa}
\email{e.wakakuwa@gmail.com}
\affiliation{Department of Communication Engineering and Informatics, Graduate School of Informatics and Engineering, The University of Electro-Communications, Japan}%


 \begin{abstract}
We propose the {\it gentle measurement principle (GMP)} as one of the principles at the foundation of quantum mechanics.
It asserts that if a set of states can be distinguished with high probability, they can be distinguished by a measurement that leaves the states almost invariant, including correlation with a reference system.
While GMP is satisfied in both classical and quantum theories, 
we show, within the framework of general probabilistic theories, that it imposes strong restrictions on the law of physics.
First, the measurement uncertainty of a pair of observables cannot be significantly larger than the preparation uncertainty.
Consequently, the strength of the CHSH nonlocality cannot be maximal.
The parameter in the {\it stretched quantum theory}, a family of general probabilistic theories that includes the quantum theory, is also limited.
Second, the conditional entropy defined in terms of a data compression theorem satisfies the chain inequality.
Not only does it imply information causality and Tsirelson's bound, but it singles out the quantum theory from the stretched one.
All these results show that GMP would be one of the principles at the heart of quantum mechanics.
\end{abstract}  

\maketitle

\prlsection{Introduction}
One of the most fundamental tasks in quantum information processing is state discrimination \cite{chefles2000quantum,barnett2009quantum,bae2015quantum,bergou2007quantum},
in which one aims at distinguishing a set of states as precisely as possible.
It is at the basis of various tasks in quantum information, such as quantum communication and quantum cryptography.
A notable property of quantum state discrimination is that if a set of states can be distinguished with high probability, they can, in principle, be distinguished by a measurement that disturbs the states only slightly.
This property is implicitly used, e.g., in the proof of a quantum capacity theorem for simultaneous transmission of classical and quantum information \cite{devetak2005capacity}.
In fact, however, this property is highly nontrivial from an operational viewpoint, given that a quantum measurement generally causes disturbance to the states to be measured \cite{Heisenberg:1927aa}.

  \begin{figure}[t]
\begin{center}
\includegraphics[bb={0 40 625 234}, scale=0.395]{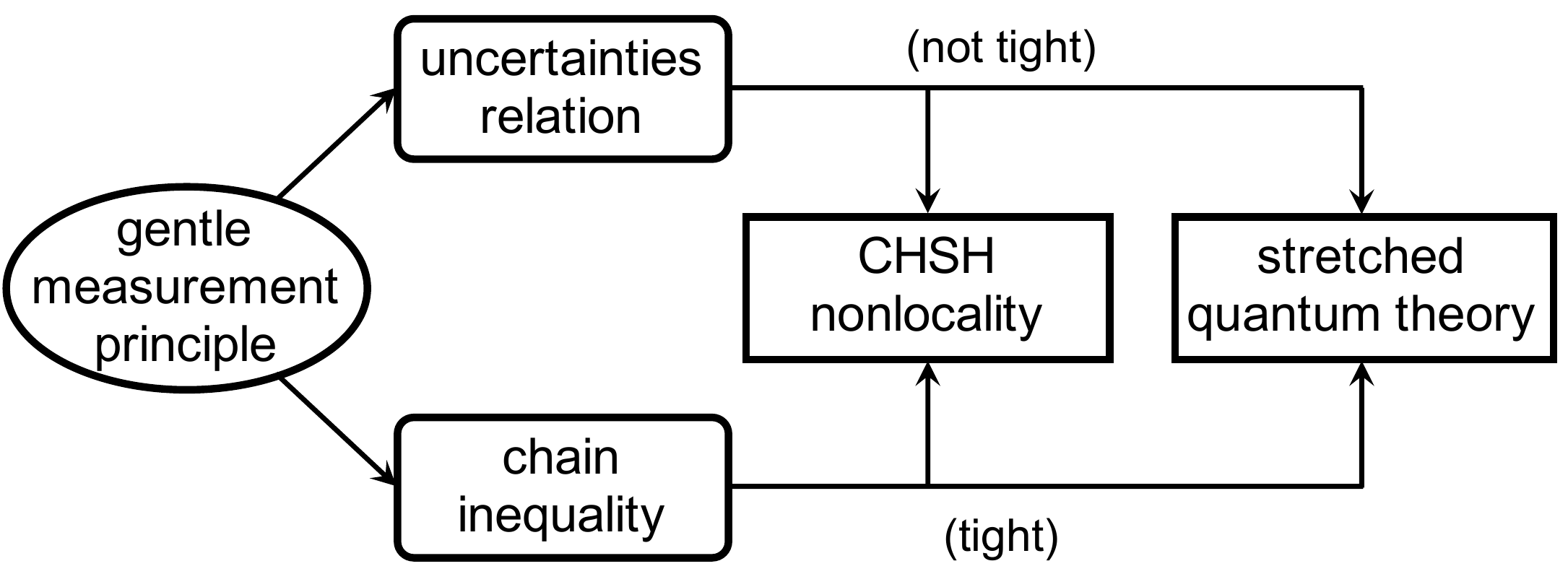}
\end{center}
\caption{
The approaches in this paper are summarized.
}
\label{fig:flowchart}
\end{figure}

In this paper, we investigate roles of the above property in the foundation of quantum theory.
For this purpose, 
we formulate the {\it gentle measurement principle (GMP)} within the framework of general probabilistic theories: 
if a set of states can be distinguished with high probability, they can be distinguished by a measurement that leaves the states almost invariant, including correlation with a reference system.
It is satisfied in both classical and quantum theories, but not necessarily in general probabilistic theories.
Indeed, GMP imposes strong restrictions on the law of physics (Fig.~\ref{fig:flowchart}).
Based on GMP, we prove that measurement uncertainty of a pair of observables cannot be significantly larger than preparation uncertainty, and that the conditional entropy defined by a data compression theorem satisfies the chain inequality.

We apply the results to investigate nonlocality in the CHSH scenario \cite{clauser1969proposed} and the {\it stretched quantum theory} (SQT).
SQT is a family of general probabilistic theories characterized by one real parameter and includes the quantum theory as a particular case.
First, by using the relation between the uncertainties, we bound the strength of the CHSH nonlocality and the value of the parameter of SQT, proving that they are limited.
However, the bounds are not tight as they do not coincide with the quantum mechanical limits.
We next apply the chain inequality with the aid of the analysis of information causality \cite{pawlowski09}.
We show that the chain inequality implies precisely the quantum mechanical limit on the CHSH nonlocality (Tsirelson's bound \cite{cirel1980quantum}) and singles out the exact quantum theory from SQT.
These results indicate that the gentle measurement principle could be one of the principles at the heart of quantum mechanics.

\prlsection{General Probabilistic Theories}
We present the postulates of general probabilistic theories that are used in this paper (see e.g.~\cite{chiribella2010probabilistic,chiribella2011informational,hardy2001quantum,barrett2007information,barnum2007generalized}).
A physical system $A$ is equipped with a set of {\it states} $\mfk{S}(A)$ and a set of {\it measurements} $\mfk{M}(A)$.
For simplicity, we assume that the set of the outcomes of a measurement $\mu\in\mfk{M}(A)$, which we denote by $\mfk{R}(\mu)$, is finite.
The probability of obtaining an outcome $r\in\mfk{R}(\mu)$ when one perfoms a measurement $\mu\in\mfk{M}(A)$ on a state $\phi\in\mfk{S}(A)$ is given by $p(r|\phi,\mu)$.
For a pair of systems $A$ and $B$, there exists a set of {\it operations} $\mfk{O}(A\rightarrow B)$.
An operation $\ca{E}\in\mfk{O}(A\rightarrow B)$ transforms a state $\phi\in\mfk{S}(A)$ to $\ca{E}(\phi)\in\mfk{S}(B)$.
The set of states is closed under probabilistic mixture,
i.e., for any $\phi_0,\phi_1\in\mfk{S}(A)$ and $q\in[0,1]$, there exists a state $\phi_q:=(1-q)\phi_0+q\phi_1\in\mfk{S}(A)$.
It holds that $p(r|\phi_q,\mu)=(1-q)p(r|\phi_0,\mu)+qp(r|\phi_1,\mu)$ for any measurement $\mu\in\mfk{M}(A)$ and any outcome $r\in\mfk{R}(\mu)$, and $\ca{E}(\phi_q)=(1-q)\ca{E}(\phi_0)+q\ca{E}(\phi_1)$ for any operation $\ca{E}\in\mfk{O}(A\rightarrow B)$.
The sequential composition of an operation $\ca{E}\in\mfk{O}(A\rightarrow B)$ and a measurement $\nu\in\mfk{M}(B)$ is a measurement on $A$.
The composition of a measurement and classical post-processing is also a measurement.

A system composed of two subsystems $A$ and $B$, which we denote by $AB$, has its own sets of states and measurements.
For any pair of states $\phi\in\mfk{S}(A)$ and $\psi\in\mfk{S}(B)$, there exists a product state $\phi\times\psi\in\mfk{S}(AB)$.
For any pair of measurements $\mu\in\mfk{M}(A)$ and $\nu\in\mfk{M}(B)$, there exists a product measurement $\mu\times\nu\in\mfk{M}(AB)$.
Product states and product measurements follows the condition of statistical independence, i.e., it holds that $p(r,s|\phi\times\psi,\mu\times\nu)=p(r|\phi,\mu)\!\cdot p(s|\psi,\nu)$. 
Similarly, for any operations $\ca{E}\in\mathfrak{O}(A\!\rightarrow\! C)$ and $\ca{F}\in\mathfrak{O}(B\!\rightarrow\! D)$, there exists a product operation $\ca{E}\!\times\! \ca{F}\in\mathfrak{O}(AB\!\rightarrow\! CD)$ and it holds that $\ca{E}\times \ca{F}(\phi\times \psi)=\ca{E}(\phi)\times\ca{F}(\psi)$.
We assume that composition is associative, i.e., $(AB)C\!=\!A(BC)$ and $\phi\times (\psi\times \xi)\!=\!(\phi\times \psi) \times \xi$. 
We may denote a system composed of $n$ duplicates of system $A$ as
$A^n\!=\!A_1A_2\!\cdots\! A_n$ and a product state thereon by $\phi_1\times\cdots\times\phi_n$.

We assume the no-signalling condition, that no measurement on a system instantaneously affects the state of the other system.
Consider a state $\phi\in\mfk{S}(AB)$ and a product measurement $\mu\times\nu$.
The condition is equivalent to the existence of the ``reduced state'' $\phi^A\in\mfk{S}(A)$ such that the marginal distribution $\sum_{s\in\mathfrak{R}(\nu)}p(r,s|\phi,\mu\times \nu)$ is equal to $p(r|\phi^A,\mu)$ and does not depend on the choice of $\nu$.
A conditional state $\phi_{r|\mu}^B\in\mathfrak{S}(B)$ is prepared on system $B$ by performing a measurement $\mu\in\mfk{M}(A)$ and post-selecting an outcome $r\in\mathfrak{R}(\mu)$.
The probability of obtaining outcomes under a product measurement is given by $p(r,s|\phi,\mu\times \nu)=p(r|\phi^A,\mu)\cdot p(s|\phi_{r|\mu}^B,\nu)$ for any $\nu\in \mathfrak{M}(B)$ and $s\in\mathfrak{R}(\nu)$.
We assume that a measurement $\mu\in\mfk{M}(A)$ followed by a measurement $\nu_r\in\mfk{M}(B)$ depending on the outcome $r\in\mfk{R}(\mu)$ is a measurement on $AB$, which is called a sequential measurement.

A {\it classical system} is a particular type of systems, such as a coin or a dice.
It is natural to assume that a classical system $X$ is represented by a finite alphabet $\ca{X}$, a set of states $\{e_x\}_{x\in\ca{X}}$ and a measurement $\mu_X$ such that $p(x'|e_x,\mu_X)=\delta_{x,x'}$.
Any measurement on $X$ is a composition of $\mu_X$ and classical post-processing.
In addition, any state $\rho\in\mfk{S}(XA)$, with $A$ being a general probabilistic system, is decomposed as
$\rho=\sum_{x\in{\ca X}}p(x)\cdot e_x\times \phi_x$ by a probability distribution $\{p(x)\}_{x\in{\ca X}}$ and a set of states $\{\phi_x\}_{x\in{\ca X}}$ on $A$.
One can prove that a measurement on system $XA$ is described as a sequential measurement from $X$ to $A$ \cite{SUPMAT}.
A measurement process on $A$ is represented by an operation from $A$ to $A'M$, where $A'$ is the output system and $M$ is the classical system to which the measurement result is recorded.

Distinguishability of two states is measured by the Kolmogorov distance \cite{kimura2010distinguishability}, defined by
\alg{
d(\phi,\psi)
:=
\sup_{\mu\in\mathfrak{M}(A)}\frac{1}{2}\sum_{r\in\mathfrak{R}(\mu)}\!\left|p(r|\phi,\mu)-p(r|\psi,\mu)\right|.
\laeq{Koldist}
}
The Kolmogorov distance is a generalization of the trace distance in quantum theory \cite{nielsentext} and satisfies the conditions for a metric.
In particular, it satisfies the triangle inequality and the monotonicity under operations, i.e., $d(\phi,\!\psi)\leq d(\phi,\omega)+d(\omega,\!\psi)$  and $d(\phi,\!\psi)\geq d(\ca{E}(\phi),\ca{E}(\psi))$ for any states $\phi,\psi,\omega\in\mfk{S}(A)$ and operation $\ca{E}\in\mfk{O}(A\!\rightarrow\! B)$.

  \begin{figure}[t]
\begin{center}
\includegraphics[bb={0 25 267 94}, scale=0.65]{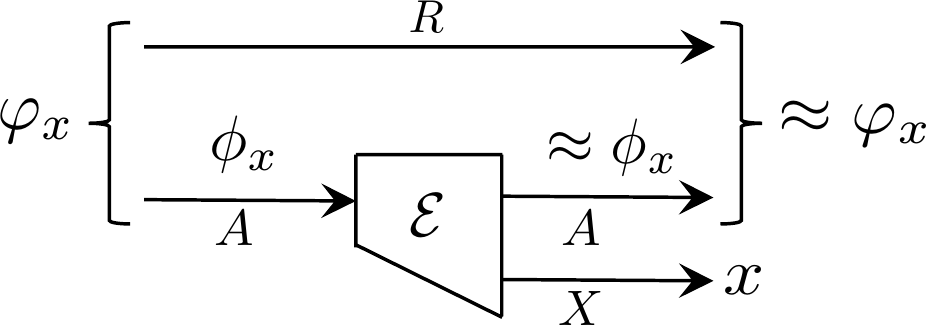}
\end{center}
\caption{
The gentle measurement principle is depicted.
}
\label{fig:GMP}
\end{figure}

\prlsection{Gentle Measurement Principle}
For formulating the gentle measurement principle (GMP), we consider a set of states $\{\phi_x\}_{x\in\ca{X}}$ on a system $A$.
Suppose that the states can be distinguished with high probability by a measurement on $A$.
GMP states that, in this case, there exists an operation on $A$ that extracts the measurement result to a classical system $X$ and leaves the states almost invariant (Fig.~\ref{fig:GMP}).
The states may have correlation with an external reference system $R$.
GMP requires that the operation also keeps the correlation almost invariant.
More precisely, suppose that there exists a measurement $\mu\in\mfk{M}(A)$ such that $\mfk{R}(\mu)=\ca{X}$ and
\alg{
p(x|\phi_x,\mu)
\geq
1-\epsilon_x
\laeq{GMP1}
}
for all $x\in\ca{X}$, where $\epsilon_x\in(0,1]$.
GMP is the principle that there exists an operation $\ca{E}:A\rightarrow AX$ and satisfies
\alg{
d(\ca{E}(\varphi_x),\varphi_x\times e_x)
\leq
\eta(\epsilon_x)
\laeq{GMP2}
}
for all $x\in\ca{X}$ and all states $\varphi_x\in\mfk{S}(AR)$ such that the reduced state on $A$ is equal to $\phi_x$.
Here, $\eta$ is a nonnegative and strictly increasing function satisfying $\lim_{\epsilon\rightarrow0}\eta(\epsilon)=0$.
GMP holds in quantum theory due to the gentle measurement lemma \cite{winter1999coding,winter1999codingth,ogawa2007making}, in which case $\eta(\epsilon)=\sqrt{\epsilon}+\epsilon/2$.

\prlsection{Uncertainties Relation}
Uncertainty relation is arguably one of the central concepts in quantum mechanics \cite{Heisenberg:1927aa}.
There are two types of uncertainty relation.
Preparation uncertainty states that no state can have definite values for both of a pair of incompatible observables \cite{Kennard:1927aa,robertson1929uncertainty}.
Measurement uncertainty states that it is impossible to simultaneously measure a pair of incompatible observables precisely \cite{arthurs1988quantum}.
In quantum theory, both of the uncertainty relations arise from the noncommutativity of operators representing the observables.
Thus, if the preparation uncertainty is strictly positive for a pair of observables, so is the measurement uncertainty.
We will refer to this property as the {\it uncertainties relation}.
Such a coincidence, however, does not necessarily hold in general probabilistic theories.
In the following, we prove that the uncertainties relation follows from GMP.

Consider a pair of measurements $\mu,\nu\in\mfk{M}(A)$ and suppose that the preparation uncertainty of them is $\epsilon$.
I.e., suppose that for any pair of measurement results $x\in\mfk{R}(\mu)$ and $y\in\mfk{R}(\nu)$, there exists a state $\phi_{xy}\in\mfk{S}(A)$ such that
\alg{
p(x|\phi_{xy},\mu)
\geq
1-\epsilon,
\quad
p(y|\phi_{xy},\nu)
\geq
1-\epsilon.
\laeq{cry}
}
Based on GMP, we can show that the minimum total error for simultaneous measurement of $\mu$ and $\nu$ cannot be significantly large. 
That is, there exists a measurement $\xi\in\mfk{M}(A)$ such that $\mfk{R}(\xi)\!=\!\mfk{R}(\mu)\!\times\!\mfk{R}(\nu)$ and it holds that
\alg{
p((x,y)|\phi_{xy},\xi)
\geq
1-2\eta(\epsilon)
\laeq{cry2}
}
for all $x$ and $y$.
Given that the total uncertainty is the sum of the two types of uncertainties,
it follows that the measurement uncertainty is small.

To prove \req{cry2}, recall that GMP and \req{cry} imply the existence of operations $\ca{E}:A\rightarrow AX$ and $\ca{F}:A\rightarrow AY$ such that
$
d(\ca{E}(\phi_{xy}),\phi_{xy}\times e_x)
\leq
\eta(\epsilon)
$
and
$
d(\ca{F}(\phi_{xy}),\phi_{xy}\times e_y)
\leq
\eta(\epsilon)
$ for all $x$ and $y$.
Due to the triangle inequality and the monotonicity of the Kolmogorov distance, we have
$
d(\ca{F}\circ\ca{E}(\phi_{xy}),\phi_{xy}\times e_x\times e_y)
\leq
2\eta(\epsilon)
$.
The distance-probability relation \req{Koldist} then implies \req{cry2}, where
$\xi\in\mfk{M}(A)$ is the sequential composition of the operation $\ca{F}\circ\ca{E}$ and the measurement $\mu_X\times\mu_Y\in\mfk{M}(XY)$.

\prlsection{CHSH Nonlocality}
The strength of nonlocality in the CHSH scenario \cite{clauser1969proposed} is limited due to the uncertainties relation.
Consider a system composed of two subsystems $A$ and $B$, which we denote by $AB$, and a state $\psi_\lambda$ on $AB$, where $\lambda\in[0,1]$.
Suppose that for a particular choice of binary measurements $\mu_0,\mu_1\in\mfk{M}(A)$ and $\nu_0,\nu_1\in\mfk{M}(B)$, it holds that
\alg{
p(r,s|\psi_\lambda,\mu_i\times\nu_j)
=
\lambda p_{PR}(r,s|i,j)
+(1-\lambda)/4,
\laeq{hanare}
}
where $p_{PR}$ is the PR-box distribution \cite{popescu1994quantum} defined by
\alg{
p_{PR}(r,s|i,j)
=
\begin{cases}
1/2 & (r\oplus s=i\cdot j)\\
0 & (r\oplus s\neq i\cdot j).\\
\end{cases}
\laeq{hanare2}
}
The parameter $\lambda$ is equal to the strength of the CHSH nonlocality up to rescaling.
The classical and quantum limits are $1/2$ and $1/\sqrt{2}$, respectively, the latter referred to as Tsirelson's bound \cite{cirel1980quantum}.
Note that any correlation in the CHSH scenario can be transformed to the {\it isotropic} form \req{hanare} by local classical pre-/post- processings \cite{masanes2006general}.
A conditional state $\psi_{\lambda,r|i}$ on $B$ is obtained by performing a measurement $\mu_i$ on $A$ and post-selecting an outcome $r$.
It satisfies
\alg{
p(r|\psi_\lambda^A,\mu_i)
\cdot
p(s'|\psi_{\lambda,r|i}^B,\nu)
=
p(r,s'|\psi_\lambda,\mu_i\!\times\!\nu)
\laeq{doshite}
}
for all $\nu\in\mfk{M}(B)$ and $s'\in\mfk{R}(\nu)$.
Note that $p(r|\psi_\lambda^A,\mu_i)=1/2$.
A simple calculation using \req{hanare} and \req{hanare2} yields
$
p(s|\psi_{\lambda,r|i}^B,\nu_j)
=
\lambda\delta_{r\oplus s,ij}
+(1-\lambda)/2
$.
Writing
$
\phi_{00}
\equiv
\psi_{\lambda,0|0}
$,
$
\phi_{01}
\equiv
\psi_{\lambda,0|1}
$,
$
\phi_{10}
\equiv
\psi_{\lambda,1|1}
$ and
$
\phi_{11}
\equiv
\psi_{\lambda,1|0}
$,
it follows that
$
p(x|\phi_{xy},\nu_0)
=
p(y|\phi_{xy},\nu_1)
=
(1+\lambda)/2
$.
Thus, due to the uncertainties relation \req{cry2}, there exists a measurement $\nu_2\in\mfk{M}(B)$ such that
\alg{
p((x,y)|\phi_{xy},\nu_2)
\geq
1-2\eta(\lambda^*),
\laeq{omoida}
}
where $\lambda^*=(1-\lambda)/2$.
We also obtain from \req{doshite} that
\alg{
&
\tilde{p}(x,y|i)
:=
\sum_{r=0,1}p(r,(x,y)|\psi_\lambda,\mu_i\!\times\!\nu_2)
\nn\\
&
=
\begin{cases}
\frac{1}{2}[p((x,y)|\phi_{00},\nu_2)\!+\!p((x,y)|\phi_{11},\nu_2)] &\! (i=0),\\
\frac{1}{2}[p((x,y)|\phi_{01},\nu_2)\!+\!p((x,y)|\phi_{10},\nu_2)] &\! (i=1).
\end{cases}
\!
\laeq{ugami}
}
Due to the no-signalling condition, the above probability distribution does not depend on $i$, that is,
$\tilde{p}(x,y|i=0)=\tilde{p}(x,y|i=1)$ for all $(x,y)$.
In particular, it must hold that
\alg{
\tilde{p}(0,0|0)
+
\tilde{p}(1,1|0)
=
1-
[\tilde{p}(0,1|1)
+
\tilde{p}(1,0|1)].
\laeq{dodo}
}
Using \req{omoida} and \req{ugami}, the L.H.S. is bounded below by
$
[p((0,0)|\phi_{00},\nu_2)+p((1,1)|\phi_{11},\nu_2)]/2
\geq
1-2\eta(\lambda^*)
$.
Similarly, the R.H.S. of \req{dodo} is bounded above by $2\eta(\lambda^*)$.
Hence, we obtain $4\eta(\lambda^*)\geq1$ and consequently arrive at
$
\lambda
\leq
1-2\eta^{-1}(1/4)
$.
Thus, the CHSH nonlocality cannot be maximal in any theory satisfying GMP.

Physically, one can interpret the above derivation as follows:
Suppose that the strength of the CHSH nonlocality is close to the maximal, i.e., $\lambda\approx1$.
The preparation uncertainty of the pair of measurements $(\nu_0,\nu_1)$ on system $B$ is then close to zero because the preparation uncertainty limits the strength of nonlocality \cite{oppenheim2010uncertainty}.
The uncertainties relation implies that the measurement uncertainty of $(\nu_0,\nu_1)$ is also close to zero.
In that case, however, Bob can simultaneously perform the two measurements within a small error and obtain information about the choice of the measurement by Alice, leading to violation of the no-signalling condition.
Thus, by contradiction, the CHSH nonlocality cannot be close to the maximal.
We remark that, although Ref.~\cite{oppenheim2010uncertainty} proved that the preparation uncertainty limits nonlocality, it did not explain why the measurements must be preparation uncertain.
Our derivation shows that it is due to GMP and the no-signalling condition.

\prlsection{Stretched Quantum Theory}
We introduce the {\it stretched quantum theory} (SQT), a family of general probabilistic theories that includes the quantum theory as a particular case.
It has one real number $\tau\in[0,1]$ as the stretching parameter and coincides with the quantum theory exactly when $\tau=0$.
Our construction of SQT is along the same line as that of the approximate quantum theory \cite{yoshida2020perfect},  
which defines the sets of states and measurements in terms of dual cones.

We assume that a physical system is represented by a finite-dimensional Hilbert space $\ca{H}$ with a fixed orthonormal basis $\{\ket{c_i}\}_{i=1}^d$ and the Fourier basis $\{\ket{f_i}\}_{i=1}^{d}$, where $d=\dim{\ca{H}}$.
Consider the set of Hermitian operators defined by
$
\mfk{S}'(A)
\!:=\!
\{
\rho\!\in\!\ca{L}(\ca{H})
\:|\:
\rho\!=\!\rho^\dagger,{\rm Tr}[\rho]\!=\!1,
\forall i;\langle c_i|\rho|c_i\rangle\!\geq\!0,\langle f_i|\rho|f_i\rangle\!\geq\!0
\}
$,
and let the set of states be $\mfk{S}(A)\!:=\!\{\rho\in\mfk{S}'(A)\:|\:\rho\!\geq\!-\tau\theta I\}$, where $\theta=(\sqrt{2}-1)/2$.
For a multipartite system $\bar{A}$ consisting of the subsystems $A_1,\cdots,A_N$, we define
$
\mfk{S}(\bar{A})
:=
\{
\sum_{k=1}^Kp_k\rho_k^{A_1}\otm\cdots\otm\rho_k^{A_N}
|
K\in\mbb{N},\rho_k^{A_j}\in\mfk{S}(A_j),
p_k\geq0, \sum_{k=1}^Kp_k=1
\}.
$
The dual cone of the set $\mfk{S}(\bar{A})$ is given by
$
\mfk{S}^*(\bar{A})
:=
\{X\in\ca{L}(\ca{H})|\forall Y\in\mfk{S}(\bar{A});\:{\rm Tr}[XY]\geq0\}
$.
We consider the positive semidefinite subset of $\mfk{S}^*(\bar{A})$, i.e.,
$
\mfk{S}_+^*(\bar{A})
:=
\{X\in\mfk{S}^*(\bar{A})|X\geq0\}
$.
The dual cone of it is given by
$
\mfk{S}_+^{**}(\bar{A})
:=
\{X\in\ca{L}(\ca{H})|\forall Y\in\mfk{S}^*(\bar{A});\:{\rm Tr}[XY]\geq0\}
$.
We define the set of states and that of measurements by
$
\ca{S}(\bar{A})
:=\{\rho\in\mfk{S}_+^{**}(\bar{A})|{\rm Tr}[\rho]=1\}
$
and
$
\ca{M}(\bar{A})
:=\{\{M_r\}_r|M_r\in\mfk{S}_+^{*}(\bar{A}),\sum_r M_r=I\}
$, 
respectively, where $I$ is the identity operator.
The probability of obtaining an outcome $r$ when one performs a measurement $\{M_r\}_r$ on a state $\rho$ is $p_r={\rm Tr}[\rho M_r]$.

To prove that the uncertainties relation limits the value of $\tau$,
let $A$ be a two-dimensional system and consider the states
$
\rho_{kl}
=(q(\tau)/\sqrt{2})[(-1)^k\sigma_z+(-1)^l\sigma_x]+I/2
$ ($k,l=0,1$),
where $\sigma_z$ and $\sigma_x$ are the Pauli $z$- and $x$- operators and $q(\tau)=(1-\tau)/2+\tau/\sqrt{2}$.
It holds that $\rho_{kl}\geq-\tau\theta I$.
Consider the measurement on $A$ with respect to the $z$-basis $\{\ket{k_z}\}_{k=0,1}=\{\ket{0},\ket{1}\}$ and the one in terms of the $x$-basis $\{\ket{l_z}\}_{l=0,1}=\{\ket{+},\ket{-}\}$.
A simple calculation yields
$
\bra{k_z}\rho_{kl}\ket{k_z}
=
\bra{l_x}\rho_{kl}\ket{l_x}
=
1-\epsilon(\tau)
$,
where $\epsilon(\tau):=(2-\sqrt{2})(1-\tau)/4$.
Thus, due to the uncertainties relation, there exists a measurement $\xi\in\mfk{M}(A)$ such that
$
p(k,l|\rho_{kl},\xi)
\geq
1-2\eta(\epsilon(\tau))
$.
Let $K$ and $L$ be random variables that take values in $\{0,1\}$ with the uniform distribution and are encoded to the state $\rho_{kl}$.
Let $\hat{K}$ and $\hat{L}$ be the results of the measurement $\xi$.
The classical mutual information between $(K,L)$ and $(\hat{K},\hat{L})$ is bounded below by Fano's inequality \cite{cover05} as $I_C(K,L\!:\!\hat{K},\hat{L})\geq2-h(2\eta(\epsilon(\tau)))-4\eta(\epsilon(\tau))$,
where $h$ is the binary entropy defined by $h(x):=-x\log{x}-(1-x)\log{(1-x)}$.
One can also prove that 
$
I_C(K,L\!:\!\hat{K},\hat{L})_\rho\leq1
$ \cite{SUPMAT}. 
Thus, a simple calculation \cite{bientineq} leads to $\tau\leq1-(4+2\sqrt{2})\eta^{-1}((1-2\ln{2})/4)<1$.

\prlsection{Chain Inequality}
In classical and quantum Shannon theories, the conditional entropy quantifies the minimum amount of classical communication required for transmitting the complete information about a random variable to the receiver in the presence of side information \cite{slepian71,devetak2003classical} (see however \cite{horo05,horo07} for the fully quantum scenario).
Based on this fact, we define the conditional entropy in general probabilistic theories.

Let $XA$ be a system composed of a classical system $X$ and a general probabilistic system $A$.
The system $X$ is in a state labeled by $x\!\in\!\ca{X}$ with probability $p(x)$, and $A$ is in the state $\phi_x$ correspondingly.
We consider a task in which the sender, who has access to $X$, sends a classical message $M$ depending on $X$ to the receiver, who subsequently performs a measurement on $A$ depending on $M$, to recover $X$.
With $n$ and $R$ denoting the block length and the communication rate, a protocol is represented by an encoding function $f\!:\!\ca{X}^n\!\!\rightarrow\![2^{nR}]$ and a set of decoding measurements $\{\mu_{m}\}_{m=1}^{2^{nR}}$.
A rate $R$ is {\it achievable} if the average decoding error can be made arbitrarily small for any sufficiently large $n$.
We define the conditional entropy $H(X|A)$ as the minimum achievable rate $R$.
In the case where the classical data to be compressed, e.g.~$X$ and $Y$, are distributed over distant parties, 
the encoding operation is performed individually on each of them by functions $f\!:\!\ca{X}^n\!\!\rightarrow\![2^{nR_X}]$ and $g\!:\!\!\ca{Y}^n\!\rightarrow\![2^{nR_Y}]$.
The conditional entropy $H(X,\!Y|A)$ is defined as the minimum of the total communication rate $R_X\!+\!R_Y$ that is achievable in this scenario.
By definition, the conditional entropies are monotonically non-decreasing under local operations on $A$ alone. 
We define the mutual information by $I(X\!:\!A)\!:=\!H(X)\!-\!H(X|A)$ and $I(X,\!Y\!:\!A)\!:=\!H(X,\!Y)\!-\!H(X,\!Y|A)$, where $H(X)$ and $H(X,\!Y)$ are the Shannon entropy.

Under the assumption of GMP, the conditional entropy satisfies the chain inequality, i.e., 
\alg{
H(X,Y|A)\leq H(X|A)+H(Y|AX).
\laeq{ci}
}
To prove this, consider a protocol for compressing $(X,Y)$ as follows.
In the first step, $X$ is compressed at rate $R_X\approx H(X|A)$ with $A$ being the side information.
Since a measurement on $A^n$ can decode $X^n$ within a small error, there exists an operation that extracts $X^n$ from $A^n$ almost perfectly while keeping the state on $A^nY^n$ almost invariant.
In the second step, $Y$ is compressed at rate $R_Y\approx H(Y|AX)$, with $AX$ serving as the side information.
The protocol achieves the total communication rate $R_X+R_Y\approx H(X|A)+H(Y|AX)$, which yields \req{ci}.
Note that the proof does not rely on the form of $\eta$ (see \cite{SUPMAT} for the detail).
The chain inequality of the mutual information immediately follows:
\alg{
I(X,Y:A)
+
I(X:Y)
\geq
I(X:A)
+I(Y:AX).
\laeq{CI}
}
Note that the equality holds in classical and quantum theories \cite{cover05,nielsentext}.

\prlsection{Implications of The Chain Inequality}
Information causality (IC) is the principle that the efficiency of nonlocality-assisted random access coding cannot be greater than the bit length of the classical message communicated in a protocol \cite{pawlowski09}.
IC does not hold in any theory in which the strength of the CHSH nonlocality can be strictly larger than the quantum mechanical limit called Tsirelson's bound \cite{cirel1980quantum}.
On the other hand, IC holds in any no-signalling theory in which one can define a ``mutual information'' satisfying the following properties:
(i) {\it Nonnegativity}: $I(X\!:\!A)\geq0$.
(ii) {\it Consistency}: If $A$ is a classical system, $I(X\!:\!A)$ coincides with the classical mutual information.
(iii) {\it Data Processing Inequality}: Under any local transformation that maps the states
of system $A$ into the states of another system $A'$ without post-selection, $I(X\!:\!A)\geq I(X\!:\!A')$.
(iv) {\it Chain Rule}: $I(X,Y\!:\!A)
+
I(X\!:\!Y)
=
I(X\!:\!A)
+I(Y\!:\!AX)$.
The mutual information defined in the previous section satisfies the properties (i), (ii), and (iii).
From the proof in \cite{pawlowski09}, we observe that one can relax the property (iv) to the chain inequality \req{CI}.
Hence, we obtain a series of implications:
GMP $\Rightarrow$ Chain Inequality $\Rightarrow$ IC $\Rightarrow$ Tsirelson's bound.
Note that, due to the same reasoning, GMP implies the information content principle \cite{czekaj2017information} as well.
Note also that Ref.~\cite{wakakuwa2012chain} derived IC from the chain rule of the generalized mutual information defined in terms of a channel coding theorem.

Next we apply the chain inequality to analyze the stretched quantum theory (SQT).
In \cite{SUPMAT}, we prove that the chain inequality \req{CI} holds if and only if the stretching parameter $\tau=0$ in SQT.
Thus, the chain inequality of the mutual information, and consequently GMP, singles out the exact quantum theory from SQT.
To prove this, we consider a protocol in which an array $X^N=X_1\cdots X_N$ of completely random bits of length $N=2^n$ is encoded into an $(N-1)$-qubit system $A^{N-1}$ with the assistance of one classical bit $M$, in such a way that $X^N$ and $A^{N-1}$ are uncorrelated unless $M$ is given.
One can prove that $I(X_0,\cdots\!,X_{N-1}\!:\!MA^{N-1})\leq1$.
From the assumption that the mutual information satisfies the chain inequality \req{CI}, we also obtain
$
I(X_0,\cdots\!,X_{N-1}\!:\!MA^{N-1})
\geq
\sum_{i=1}^N
I(X_i\!:\!MA^{N-1}X_0\cdots X_{i-1})
\geq
\sum_{i=1}^N
I(X_i\!:\!MA^{N-1})
\geq
\sum_{i=1}^N[1\!-\!h(P_i)]
=:
J_n$.
Here, $P_i$ is the probability of correctly guessing $X_i$ by a measurement on $MA^{N-1}$.
Thus, if a protocol achieves $J_n>1$, the chain inequality does not hold.
We prove in \cite{SUPMAT} that such a protocol exists for any $\tau>0$ and sufficiently large $n$.

\prlsection{Conclusion}
In this paper, we have proposed the gentle measurement principle (GMP) as one of the principles at the foundation of quantum mechanics.
GMP implies the uncertainties relation and the chain inequality of the conditional entropy.
Based on these results, we analyzed the CHSH nonlocality and the stretched quantum theory, proving that GMP imposes strong restrictions on the law of physics. 
In particular, the chain inequality implies Tsirelson's bound for the CHSH nonlocality and singles out the exact quantum theory from the stretched one. 
The concept of GMP is in some aspect similar to that of measurement sharpness \cite{chiribella2014measurement} but is different: the disturbance caused by a measurement is formulated in terms of states in GMP, not of measurements as in \cite{chiribella2014measurement}, and GMP incorporates cases where disturbance is not exactly zero.
A future direction is to investigate relations of GMP with other concepts in quantum foundations, 
such as purification \cite{chiribella2010probabilistic}, local orthogonality \cite{fritz2013local} and the existence of an information unit \cite{masanes2013existence}.

\prlsection{Acknowledgement}
This work is supported by JSPS KAKENHI (Grant No.~18J01329), Japan.

\bibliography{bibbib.bib}

\newpage

\setcounter{page}{1}

\appendix

\section*{Supplemental Material}

This material is organized as follows.
In \rApp{GMPQ}, we prove that the gentle measurement principle holds in quantum theory.
In \rApp{adas}, we introduce some additional assumptions in general probabilistic theories and present a few lemmas that we use to obtain the main results.
The proofs of the lemmas will be provided at the end of this material, \rApp{prfs}.
In \rApp{dfnCE}, we present a rigorous definition of conditional entropy and describe its properties.
\rApp{PrfCE} provides detailed proof of the chain inequality of the conditional entropy under the assumption of the gentle measurement principle.
\rApp{prfineq} and \rapp{nestVD} are devoted to investigating protocols in the stretched quantum theory.
In \rApp{prfineq}, we prove that the amount of classical information encoded to a qubit system by a protocol cannot be greater than $1$.
In \rApp{nestVD}, we introduce a variant of the {\it nested van Dam's protocol} \cite{pawlowski09} and prove that the information gain can be strictly greater than $1$ if $\tau>0$. 
A refinement of the uncertainties relation is presented in \rApp{extUR}.

\section{Gentle Measurement Lemma}
\lapp{GMPQ}

We prove that the gentle measurement principle holds in quantum theory.
We denote the set of linear operators on a finite-dimensional Hilbert space $\ca{H}$ by $\ca{L}(\ca{H})$, and the set of normalized states thereon by $\ca{S}(\ca{H})$, i.e.,
\alg{
\ca{S}(\ca{H})
:=
\{\rho\in\ca{L}(\ca{H})|\rho\geq0,{\rm Tr}[\rho]=1\}.
}
The proof is based on the gentle measurement lemma, which was originally proved in \cite{winter1999coding,winter1999codingth} and was later improved in \cite{ogawa2007making}:

\blmm{GML} {\bf (Lemma 5 in \cite{ogawa2007making})}
Let $\ca{H}$ be a finite-dimensional Hilbert space.
For any state $\rho\in\ca{S}(\ca{H})$ and any operator $X\in\ca{L}(\ca{H})$ such that $0\leq X\leq I$, it holds that
\alg{
\left\|\rho-\sqrt{X}\rho\sqrt{X}\right\|_1
\leq
2\sqrt{1-{\rm Tr}[\rho X]},
}
where $\|\cdot\|_1$ is the trace norm defined by $\|Y\|_1:={\rm Tr}[\sqrt{Y^\dagger Y}]$ for $Y\in\ca{L}(\ca{H})$.
\elmm

To prove that the quantum theory satisfies the gentle measurement principle, consider a set of states $\{\phi_x\}_{x\in\ca{X}}$ on a finite-dimensional quantum system $A$.
Suppose that the states can be distinguished within an error $\epsilon_x\in(0,1]$ for each $x$.
That is, suppose that there exists a measurement on $A$, represented by a positive operator-valued measure (POVM) $\{M_x\}_{x\in\ca{X}}$, such that
\alg{
{\rm Tr}[M_x\phi_x]\geq1-\epsilon_x.
\laeq{juge}
}
Let $X$ be a ``classical'' system with a fixed orthonormal basis $\{|x\rangle\}_{x\in\ca{X}}$.
Define an operation (completely positive trace-preserving map) $\ca{E}:A\rightarrow AX$ by
\alg{
\ca{E}(\cdot)
:=
\sum_{x\in\ca{X}}\sqrt{M_x}(\cdot)\sqrt{M_x}^A\otm\proj{x}^X.
}
With $R$ denoting a reference system,
we prove that for any $x\in\ca{X}$ and for any state $\varphi_x\in\ca{S}(\ca{H}^A\otm\ca{H}^R)$ satisfying ${\rm Tr}_R[\varphi_x]=\phi_x$, it holds that
\alg{
\!
\frac{1}{2}
\left\|\ca{E}\otm{\rm id}^R(\varphi_x)-\varphi_x\otm\proj{x}^X\right\|_1
\leq
\sqrt{\epsilon_x}+\frac{\epsilon_x}{2}.
\!
}

Due to \req{juge}, we have
\alg{
{\rm Tr}[(M_x^A\otm I^R)\varphi_x]\geq1-\epsilon_x.
\laeq{juge2}
}
Noting that $\sqrt{M_x\otm I}=\sqrt{M_x}\otm I$, the gentle measurement lemma (\rLmm{GML}) implies that
\alg{
\left\|
(\sqrt{M_x}^A\otm I^R)\varphi_x(\sqrt{M_x}^A\otm I^R)
-\varphi_x
\right\|_1
\leq
2\sqrt{\epsilon_x}.
}
From \req{juge2} and $\sum_{x\in\ca{X}}M_x=I$, we also have
\alg{
&
\sum_{x'(\neq x)}\left\|(\sqrt{M_{x'}}^A\otm I^R)\varphi_x(\sqrt{M_{x'}}^A\otm I^R)\right\|_1
\nn\\
&=
\sum_{x'(\neq x)}
{\rm Tr}[(M_{x'}^A\otm I^R)\varphi_x]
\\
&=
{\rm Tr}[((I-M_{x})^A\otm I^R)\varphi_x]
\leq\epsilon_x.
\laeq{juge3}
}
Hence, we obtain
\alg{
&
\left\|\ca{E}\otm{\rm id}^R(\varphi_x)-\varphi_x\otm\proj{x}^X\right\|_1
\nn\\
&
=
\left\|
(\sqrt{M_x}^A\otm I^R)\varphi_x(\sqrt{M_x}^A\otm I^R)
-\varphi_x
\right\|_1
\nn\\
&\quad
+\sum_{x'(\neq x)}\left\|(\sqrt{M_{x'}}^A\otm I^R)\varphi_x(\sqrt{M_{x'}}^A\otm I^R)\right\|_1
\\
&\leq
2\sqrt{\epsilon_x}+\epsilon_x,
}
which completes the proof.
\QED

\section{Additional Assumptions and Lemmas}
\lapp{adas}

We introduce some additional assumptions in general probabilistic theories that we use to obtain the main results.
First, we assume that there exists an identity operation ${\rm id}\in\mfk{O}(A\rightarrow A)$ such that ${\rm id}(\phi)=\phi$ for all $\phi\in\mfk{S}(A)$ and ${\rm id}^\dagger(\mu)=\mu$ for all $\mu\in\mathfrak{M}(A)$. 
For any operations $\ca{E}\in\mfk{O}(A\rightarrow B)$ and $\ca{E}'\in\mfk{O}(B'\rightarrow A)$, it holds that $\ca{E}\circ{\rm id}=\ca{E}$ and ${\rm id}\circ\ca{E}'=\ca{E}'$.
We require that ${\rm id}^A\times{\rm id}^B={\rm id}^{AB}$.
For simplicity, we denote ${\rm id}^A\times\ca{E}(\psi)$, where $\psi\in\mfk{S}(AB)$ and $\ca{E}\in\mfk{O}(B\rightarrow C)$, by $\ca{E}(\psi)$.
Second, we assume that the product states satisfy the affinity in the sense that 
$\phi_q\times\psi=(1-q )\phi_0\times\psi+q  \phi_1\times\psi$, where $\phi_q=(1-q )\phi_0+q  \phi_1$.
Finally, we require existence of certain types of operations.
For any state $\psi\in \mathfrak{S}(B)$, there exists an operation $\ca{E}_\psi\in\mathfrak{O}(A\rightarrow AB)$ such that for any $\phi \in \mathfrak{S}(A)$, it holds that
\alg{
\ca{E}_\psi(\phi)=\phi\times \psi.
\laeq{dfnjointop}
}

Consider a system $XA$ composed of a classical system $X$ and a general probabilistic system $A$, which we refer to as a {\it C-G system}.
Any measurement on $XA$ can be represented as a sequential composition of a measurement on $X$ and the subsequent measurement on $A$:

\blmm{CGmeasdec}{\bf [Measurements on C-G System.]}
For any measurement $\nu\in\mathfrak{M}(XA)$, there exists a set of measurement $\{\nu_x\}_{x\in\ca{X}}$ on $A$ such that $\mathfrak{R}(\nu_x)=\mathfrak{R}(\nu)$ and for any state $\rho\in\mathfrak{S}(XA)$, which is decomposed into
\alg{
\rho=\sum_{x\in{\ca X}}p(x)\cdot e_x\times \phi_x,
\laeq{rhodfn}
}
it holds that
\alg{
p(s|\rho,\nu)
=
\sum_{x\in\ca{X}}p(x)p(s|\phi_x,\nu_x).
\laeq{iwasok}
}
In addition, there exists a measurement $\tilde{\nu}\in\mathfrak{M}(XA)$ such that $\mathfrak{R}(\tilde{\nu})=\ca{X}\times\mathfrak{R}(\nu)$ and it holds that
\alg{
p(s|\rho,\nu)
&
=
\sum_{x\in\ca{X}}p(x,s|\rho,\tilde{\nu}),
\laeq{iwasok2}
\\
p(x',s|e_x\times \varphi,\tilde{\nu})
&=
\delta_{x,x'}\cdot p(s|\varphi,\nu_x)
\laeq{iwasok3}
}
for all $\varphi\in\mathfrak{S}(A)$.
\elmm

\noindent
The Kolmogorov distance takes a particular form for the states on $XA$:

\blmm{TDCG}{\bf [Kolmogorov Distance in C-G System.]}
Consider states $\rho,\sigma\in\mathfrak{S}(XA)$ such that
\alg{
\rho\!=\!\sum_{x\in\ca{X}}p_xe_x\!\times\! \phi_x,\:
\sigma\!=\!\sum_{x\in\ca{X}}p_xe_x\!\times\! \psi_x,
}
where $\{p_x\}_x$ and $\{q_x\}_x$ are probability distributions and $\phi_x,\psi_x\in\mfk{S}(A)$ for all $x\in\ca{X}$.
It holds that
\alg{
d(\rho,\sigma)
=
\sum_{x\in\ca{X}}p_xd(\phi_x,\psi_x).
\laeq{dRS1}
}
\elmm

\noindent
Proofs of \rLmm{CGmeasdec} and \rlmm{TDCG} are provided in \rApp{prfs}.

\section{Definitions of Conditional Entropies}
\lapp{dfnCE}

To define the conditional entropies, we consider a state $\rho$ on a system $XA$, composed of a classical system $X$ and a general probabilistic system $A$.
Due to the definition of the classical system (see the main text), the state $\rho$ is represented as
\alg{
\rho=\sum_{x\in{\ca X}}p(x)\cdot e_x\times \phi_x,
}
where $\{p(x)\}_x$ is a probability distribution, $\{e_x\}_x$ is the basis states of $X$ and $\{\phi_x\}_x$ is a set of states on $A$.
The state is equivalently represented by an ensemble $\{p(x),\phi_x\}_{x\in{\mathcal X}}$ on system $A$.
The expected state of this ensemble is given by
\alg{
\phi=\sum_{x\in{\ca X}}p(x)\cdot \phi_x,
}
which is equal to the reduced state of $\rho$ on $A$.
For $n\in\mbb{N}$, we introduce notations 
$
x^n:=x_1\cdots x_n\in\ca{X}^n
$
and
\alg{
p(x^n)
:=
p(x_1)\cdots p(x_n),
\quad
\phi^n_{x^n}:=\phi_{x_1\!}\times\cdots\times\!\phi_{x_n}.
\laeq{imafuta}
}

For the definition of conditional entropy,
we consider the task of classical data compression with general probabilistic side information.
The sender, who has access to $X$, aims at providing complete information about $X$ to the receiver, who has access to $A$.
To this end, the sender performs classical processing on $X^n$ to obtain a message $M$ and sends it to the receiver.
The receiver performs a measurement on $A^n$ depending on $M$ to obtain the information about $X^n$.
We define the conditional entropy as the minimum ratio of the length of the message required for achieving this task with vanishingly small error:

\bdfn{SSW}{\bf [Conditional Entropy.]}
A pair of a function $f:\ca{X}^n\rightarrow[2^{nR}]$ and a set of measurements $\{\mu_{x'}\}_{x'\in[2^{nR}]}$ on $A^n$ is called an $(n,2^{nR},\epsilon)$ code for the state $\rho\in\mfk{S}(XA)$ if it holds that $\mathfrak{R}(\mu_{x'})=\ca{X}^n$ and
\alg{
\sum_{x^n\in \ca{X}^n}p(x^n|\phi_{x^n}^n,\mu_{x'=f(x^n)})p(x^n)\geq1-\epsilon.
}
A rate $R$ is {\it achievable for the state $\rho\in\mfk{S}(XA)$} if there exists a sequence of $(n,2^{nR},\epsilon_n)$ codes for $\rho$ ($n=1,2,\cdots)$ such that $\lim_{n\rightarrow\infty}\epsilon_n=0$. 
The conditional entropy of $X$ conditioned by $A$ in the state $\rho$, which we denote by $H(X|A)_{\rho}$, is defined as the infimum of a rate $R$ that is achievable for the state $\rho$.
\edfn

Next, we consider a scenario in which the classical data to be compressed is distributed over two distant parties.
We consider a state $\rho$ on a system $XYA$, composed of classical systems $X$, $Y$ and a general probabilistic system $A$.
The state $\rho$ is represented as
\alg{
\rho=\sum_{x\in{\ca X},y\in{\ca Y}}p(x,y)\cdot e_x\times e_y\times \phi_{xy},
}
where $\{p(x,y)\}_{x,y}$ is a probability distribution, $\{e_x\}_x$ and $\{e_y\}_y$ are the basis states of $X$ and $Y$, and $\{\phi_{xy}\}_{x,y}$ is a set of states on $A$.
The state is equivalently represented by an ensemble $\{p(x,y),\phi_{xy}\}_{x\in{\mathcal X},y\in{\mathcal Y}}$ on system $A$.
The expected state of this ensemble is given by
\alg{
\phi=\sum_{x\in{\mathcal X},y\in{\mathcal Y}}p(x,y)\cdot \phi_{xy},
}
which is equal to the reduced state of $\rho$ on $A$.
For $n\in\mbb{N}$, we introduce notations 
$
x^n:=x_1\cdots x_n\in\ca{X}^n
$,
$
y^n:=y_1\cdots y_n\in\ca{Y}^n
$
and
\alg{
p(x^n,y^n)
&
:=
p(x_1,y_1)\cdots p(x_n,y_n),
\laeq{imafutatari}
\\
\phi^n_{x^ny^n}
&:=\phi_{x_1y_1}\times\cdots\times \phi_{x_ny_n}.
\laeq{imafutata}
}
We define the entropy of the system $X,Y$ conditioned by $A$ as follows:

\bdfn{SSWdis}{\bf [Conditional Entropy.]}
A triplet of a function $f:\ca{X}^n\rightarrow[2^{nR_X}]$, $g:\ca{Y}^n\rightarrow[2^{nR_Y}]$ and a set of measurements $\{\mu_{x'y'}\}_{x'\in[2^{nR_X}],y'\in[2^{nR_Y}]}$ on $A^n$ is called an $(n,2^{nR_X},2^{nR_Y},\epsilon)$ code for the state $\rho\in\mfk{S}(XYA)$ if it holds that $\mathfrak{R}(\mu_{x'y'})=\ca{X}^n\times\ca{Y}^n$ and that 
\begin{eqnarray}
\sum_{\substack{x^n\in \ca{X}^n\\y^n\in \ca{Y}^n}}p((x^n,y^n)|\phi_{x^ny^n}^n,\mu_{x'=f(x^n),y'=f(y^n)})p(x^n,y^n)
\nn\\
\geq1-\epsilon.
\quad
\end{eqnarray}
A rate pair $(R_X,R_Y)$ is {\it achievable for the state $\rho\in\mfk{S}(XYA)$} if there exists a sequence of $(n,2^{nR},\epsilon_n)$  codes for $\rho$ ($n=1,2,\cdots)$ such that $\lim_{n\rightarrow\infty}\epsilon_n=0$. 
The conditional entropy of the system $X,Y$ conditioned by $A$ in the state $\rho$, which we denote by $H(X,Y|A)_{\rho}$, is defined as the infimum of a rate sum $R_X+R_Y$ such that $(R_X,R_Y)$ is achievable for the state $\rho$.
\edfn

The mutual informations for non-distributed and distributed scenarios are defined by
\alg{
I(X:A)_{\rho}&:=H(X)_{\rho}-H(X|A)_{\rho},
\laeq{damashi1}\\
I(X,Y:A)_{\rho}&:=H(X,Y)_{\rho}-H(X,Y|A)_{\rho},
\laeq{damashi2}
}
respectively, where $H(X)$ and $H(X,Y)$ are the Shannon entropies.
From the definition, it is straightforward to verify that the conditional entropy is monotonically non-decreasing under operations on $A$ without post-selection.
That is, for 
any operation $\ca{E}\in\mfk{O}(A\rightarrow A')$, it holds that 
\alg{
H(X|A)_\rho
&\leq H(X|A')_{\ca{E}(\rho)},
\\
H(X,Y|A)_\rho
&\leq H(X,Y|A')_{\ca{E}(\rho)}.
}
The monotonicity of the mutual informations immediately follows:
\alg{
I(X:A)_\rho
&\geq I(X:A')_{\ca{E}(\rho)},
\\
I(X,Y:A)_\rho
&\geq I(X,Y:A')_{\ca{E}(\rho)}.
}
For a state $\rho\in\mfk{S}(XA)$, the accessible information is defined by
\alg{
I_{acc}(X:A)_\rho
:=
\sup_{\mu\in\mfk{M}(A)}I_C(X:Y),
}
where $I_C$ is the classical mutual information, $Y$ is the result of the measurement $\mu$ and the supremum is taken over all measurements on $A$.
We omit the subscript $\rho$ when it is clear from the context.

The conditional entropies defined above coincide with the classical conditional entropy if $A$ is a classical system \cite{slepian71,cover05} and the classical-quantum conditional entropy if $A$ is a quantum system \cite{devetak2003classical}.
Correspondingly, the mutual informations \req{damashi1} and \req{damashi2} also coincide with the classical mutual information or the classical-quantum mutual information (the Holevo information \cite{holevo73}).

We will use the following lemma in \rApp{prfineq} and \rApp{nestVD}:

\blmm{omoida}
Let $X$ and $T$ be classical systems, $A$ be a general probabilistic system and consider a state $\rho\in\mfk{S}(XTA)$.
Suppose that 
\alg{
I_{acc}(X:A)_\rho=0.
\laeq{anohito}
}
Then, it holds that
\alg{
I_{acc}(X:TA)_\rho
&
\leq
H(T)_\rho,
\laeq{hirokumo}
\\
I(X:TA)_\rho
&
\leq
H(T)_\rho.
\laeq{migaki}
}
\elmm

\noindent
A proof of this lemma is provided in \rApp{prfs}.

\section{Proof of The Chain Inequality}
\lapp{PrfCE}

Under the assumption of the gentle measurement principle (GMP), we prove that the conditional entropy satisfies the chain inequality:
\alg{
H(X,Y|A)
\leq
H(X|A)
+
H(Y|AX).
\laeq{chainineqsupp3}
}
Here, $X$ and $Y$ are classical systems and $A$ is a general probabilistic system.
The conditional entropies are for the state $\rho\in\mfk{S}(XYA)$ represented by
\alg{
\rho=\sum_{x\in{\ca X},y\in{\ca Y}}p(x,y)\cdot e_x\times e_y\times \phi_{xy},
}
where $\{p(x,y)\}_{x,y}$ is a probability distribution, $\{e_x\}_x$ and $\{e_y\}_y$ are the basis states of $X$ and $Y$, and $\{\phi_{xy}\}_{x,y}$ is a set of states on $A$.
For the proof, we define
\alg{
\phi_{x}
&
:=
\sum_{y\in{\ca Y}}p(y|x)\cdot  \phi_{xy}
\;\in\mfk{S}(A),
\\
\phi_{y}
&
:=
\sum_{x\in\ca{X}}p(x|y)\cdot \phi_{xy}
\;\in\mfk{S}(A),
\\
\varphi_{x}
&
:=
\sum_{y\in{\ca Y}}p(y|x)\cdot  e_y\times \phi_{xy}
\;\in\mfk{S}(YA),
\\
\varphi_{y}
&
:=
\sum_{x\in\ca{X}}p(x|y)\cdot e_x\times \phi_{xy}
\;\in\mfk{S}(XA)
}
and 
$
\hat{\eta}(x):=\min\{\eta(x),1\}
$.
In addition to \req{imafutatari} and \req{imafutata},
we introduce notations
\alg{
\varphi^n_{x^n}
&:=\varphi_{x_1}\times\cdots\times \varphi_{x_n},
\\
\varphi^n_{y^n}
&:=\varphi_{y_1}\times\cdots\times \varphi_{y_n},
\\
e^n_{x^n}
&:=e_{x_1}\times\cdots\times e_{x_n}
}
and 
\alg{
p(x^n|y^n)
:=
p(x_1|y_1)\cdots p(x_n|y_n).
}
We
fix arbitrary $R_X>H(X|A)$, $R_Y>H(Y|AX)$, $\epsilon>0$ and choose sufficiently large $n$.

By definition, there exists a function $f:\ca{X}^n\rightarrow [2^{nR_X}]$ and a set of measurements $\{\mu_{x'}\}_{x'\in[2^{nR_X}]}$ on $A^n$ such that $\mathfrak{R}(\mu_{x'})=\ca{X}^n$
and
\alg{
\sum_{x^n\in\ca{X}^n}p(x^n)
p(x^n|\phi_{x^n}^n,\mu_{x'=f(x^n)})
\geq
1-\epsilon.
}
Denoting
$
\epsilon(x^n)\equiv 1-p(x^n|\phi_{x^n}^n,\mu_{x'=f(x^n)})
$,
we have
\alg{
\sum_{x^n\in\ca{X}^n}p(x^n)
\epsilon(x^n)
\leq
\epsilon.
\laeq{season}
}
Due to GMP, there exists a set of operations $\{\ca{F}_{x'}\}_{x'\in[2^{nR_X}]}$ from $A^n$ to $A^nX^n$ such that
\alg{
d\left(
\ca{F}_{x'=f(x^n)}(\varphi_{x^n}^n),e_{x^n}^n\times \varphi_{x^n}^n
\right)
\leq
\hat{\eta}(\epsilon(x^n)).
}
Using the property of the Kolmogorov distance (see \rLmm{TDCG}), we obtain
\alg{
&\!\!\!\!\!\!
\sum_{x^n\in\ca{X}^n,y^n\in\ca{Y}^n}
\!\!\!\!\!\!
p(x^n\!,y^n)
\cdot
d\left(
\ca{F}_{x'=f(x^n)}(\phi_{x^ny^n}^n),e_{x^n}^n\times \phi_{x^ny^n}^n
\right)
\!\!\!\!\!\!\!\!
\nn
\\
&
\quad\quad
\leq
\sum_{x^n\in\ca{X}^n}p(x^n)
\hat{\eta}(\epsilon(x^n)).
\quad\quad
\laeq{morag}
}
From \req{season} and \rLmm{yoshiki} below, the R.H.S. of the above inequality is bounded above by
$
\eta(\sqrt{\epsilon})+\sqrt{\epsilon}
$.
Similarly, there exists a function $g:\ca{Y}^n\rightarrow [2^{nR_Y}]$ and a set of measurements $\{\nu_{y'}\}_{y'\in[2^{nR_Y}]}$ 
 on $A^nX^n$ such that $\mathfrak{R}(\nu_{y'})=\ca{Y}^n$
and
\alg{
\sum_{y^n\in\ca{Y}^n}
p(y^n)
[1-p(y^n|\varphi_{y^n}^n,\nu_{y'=g(y^n)})]
\leq
\epsilon.
\laeq{subsub}
}
Due to \rLmm{CGmeasdec}, there exists a set of measurements $\{\tilde{\nu}_{y'}\}_{y'\in[2^{nR_Y}]}$ on $A^nX^n$ such that $\mathfrak{R}(\tilde{\nu}_{y'})=\ca{X}^n\ca{Y}^n$ and it holds that
\alg{
&\!\!
p(y^n|\varphi_{y^n}^n,\nu_{y'=g(y^n)})
\!=\!\!\!\!
\sum_{x^n\in\ca{X}^n}
\!\!
p(x^n\!,y^n|\varphi_{y^n}^n,\tilde{\nu}_{y'=g(y^n)})
,\!\!\!
\\
&\!\!
p(x'^n,y^n|e_{x^n}^n\times \phi_{x^ny^n}^n,\tilde{\nu}_{y'=g(y^n)})
\propto
\delta_{x^n,x'^n}.
}
It follows that
\alg{
&
p(y^n|\varphi_{y^n}^n,\nu_{y'=g(y^n)})
\nn\\
&
=
\sum_{x^n\in\ca{X}^n}\!p(x^n|y^n)p(y^n|e_{x^n}^n\times \phi_{x^ny^n}^n,\tilde{\nu}_{y'=g(y^n)})
\nn\\
&
=
\sum_{x^n\in\ca{X}^n}\!p(x^n|y^n)\!\!\sum_{x'^n\in\ca{X}^n}\!p(x'^n\!,y^n|e_{x^n}^n\!\times\!\phi_{x^ny^n}^n,\tilde{\nu}_{y'=g(y^n)})\!
\nn\\
&
=
\sum_{x^n\in\ca{X}^n}\!p(x^n|y^n)p(x^n\!,y^n|e_{x^n}^n\!\times\! \phi_{x^ny^n}^n,\tilde{\nu}_{y'=g(y^n)}).
\!\!\!
}
Substituting this to \req{subsub}, we arrive at
\begin{eqnarray}
\sum_{x^n\in\ca{X}^n,y^n\in\ca{Y}^n}p(x^n,y^n)p(x^n,y^n|e_{x^n}^n\times \phi_{x^ny^n}^n,\tilde{\nu}_{y'=g(y^n)})
\nn\\
\geq
1-\epsilon.
\quad\quad
\laeq{morag2}
\end{eqnarray}

For each $x'\in[2^{nR_X}]$ and $y'\in[2^{nR_Y}]$, let $\xi_{x',y'}$ be a measurement on $A^n$ composed of an operation $\ca{F}_{x'}\in\mfk{O}(A^n\rightarrow A^nX^n)$ followed by a measurement $\tilde{\nu}_{y'}\in\mfk{M}(A^nX^n)$.
From \req{morag}, \req{morag2} and the definition of the Kolmogorov distance \req{Koldist}, we obtain
\begin{eqnarray}
\sum_{x^n\in\ca{X}^n,y^n\in\ca{Y}^n}
\!
p(x^n\!,y^n)\!\cdot\!
p(x^n\!,y^n|\phi_{x^ny^n}^n,\xi_{x'=f(x^n),y'=g(y')})
\!\!\!\!\!\!\!\!\!
\nn\\
\geq
1-\eta(\sqrt{\epsilon})-\sqrt{\epsilon}-\epsilon.
\quad
\end{eqnarray}
Since $\epsilon$ can be arbitrarily small for sufficiently large $n$,
this implies that the rate $R=R_X+R_Y$ is achievable in distributed compression of $X$ and $Y$ with respect to the state $\rho$.
Since this relation holds for any $R_X>H(X|A)$ and $R_Y>H(Y|AX)$, we obtain \req{chainineqsupp3}.
\QED

\begin{lmm}\label{lmm:yoshiki}(Lemma 35 in \cite{wakakuwa2017coding})
Let $c\in(0,\infty)$ be a constant, $f:[0,c]\rightarrow{\mathbb R}$ be a monotonically nondecreasing function that  satisfies $f(c)<\infty$, and $\{p_k\}_{k\in{\mathbb K}}$ be a probability distribution on a countable set ${\mathbb K}$.  Suppose $\epsilon_k\:(k\in{\mathbb K})$ satisfies $\epsilon_k\in[0,c]$, and $\sum_{k\in{\mathbb K}}p_k\epsilon_k\leq\epsilon$ for a given $\epsilon\in(0,c^2]$. Then we have
\begin{eqnarray}
\sum_{k\in{\mathbb K}}p_kf(\epsilon_k)\leq f(\sqrt{\epsilon})+f( c)\cdot\sqrt{\epsilon}.\label{eq:detectivewall}
\end{eqnarray}
\end{lmm}

\section{Proof of Inequality $I(K,L:\hat{K},\hat{L})\leq1$}
\lapp{prfineq}

Let $K$ and $L$ be classical systems with the basis states $\{\ket{k}\}_{k=0,1}$ and $\{\ket{l}\}_{l=0,1}$, respectively.
We denote $\ket{k}^K\ket{l}^L$ simply by $\ket{k,l}^{KL}$.
Consider the state
\alg{
\rho=
\frac{1}{4}\sum_{k,l=0,1}\proj{k,l}^{KL}\otm\rho_{kl}^A,
\laeq{dfnrho}
}
where
\alg{
\rho_{kl}
=\frac{q(\tau)}{\sqrt{2}}[(-1)^k\sigma_z+(-1)^l\sigma_x]+\frac{1}{2}I,
\laeq{hashiru}
}
where $q(\tau)=(1-\tau)/2+\tau/\sqrt{2}$.
Let $\hat{K}$ and $\hat{L}$ be the result of the measurement performed on $A$ to guess $K$ and $L$.
We prove that $I(K,L:\hat{K},\hat{L})\leq1$ for any $\tau\geq0$. 

For each $k$, $l$ and $t\in\{0,1\}$, define the state $\rho_{kl,t}:=\sigma_y^t\rho_{kl}\sigma_y^t$.
From \req{hashiru}, we have
\alg{
\rho_{kl,t}=\rho_{k\oplus t,l\oplus t},
\laeq{sstr}
}
where $\oplus$ denotes summation modulo $2$.
Thus, we obtain
\alg{
\frac{1}{2}[\rho_{kl,0}\!+\!\rho_{kl,1}]=
\frac{1}{2}[\rho_{00}\!+\!\rho_{11}]=
\frac{1}{2}[\rho_{01}\!+\!\rho_{10}]=\pi
\laeq{kyos}
}
for any $k$ and $l$, where $\pi:=I/2$.
Consider the state
\alg{
\tilde{\rho}=
\frac{1}{8}\sum_{k,l,t=0,1}\proj{k,l}^{KL}\otm\proj{t}^T\otm\rho_{kl,t}^A.
\laeq{dfnrhotilde}
}
It follows that
\alg{
\tilde{\rho}^{KLA}=
\left(\frac{1}{4}\sum_{k,l,t=0,1}\proj{k,l}^{KL}\right)\otm\pi^A.
\laeq{kinina}
}
Hence, we obtain $I_{acc}(K,L:A)_{\tilde\rho}=0$.
Due to \rLmm{omoida} below, it follows that $I_{acc}(K,L:TA)_{\tilde\rho}\leq H(T)=1$.
From \req{sstr},
we also have $I_{acc}(K,L:TA)_{\tilde\rho}\geq I_{acc}(K,L:A)_{\rho}$.
Thus, we arrive at $I_{acc}(K,L:A)_{\rho}\leq1$, which implies $I(K,L:\hat{K},\hat{L})\leq1$.
\QED

\section{Nested van Dam's Protocol}
\lapp{nestVD}

We prove that the chain inequality of the mutual information \req{CI} does not hold in the stretched quantum theory if $\tau>0$. 
To this end, we consider a protocol in which an array $X^N=X_0\cdots X_{N-1}$ of completely random bits of length $N=2^n$ is encoded into a state $\rho$ on $(N-1)$-qubit system $A^{N-1}$ with the assistance of one classical bit $M$.
The encoding scheme is such that there is no correlation between $X^N$ and $A^{N-1}$ unless $M$ is given, i.e.,
\alg{
I_{acc}(X^N:A^{N-1})=0.
\laeq{italy}
}
Let $P_i:={\rm Pr}\{\hat{X}_i=X_i\}$ denote the probability of correctly guessing $X_i$ from the result of the measurement on $MA^{N-1}$.
We prove that the protocol achieves
\alg{
P_i=
\frac{1}{2}\left[1+\left(\frac{1+(\sqrt{2}-1)\tau}{\sqrt{2}}\right)^n\right]
\laeq{supr}
}
for each $i$.
A simple calculation using the relation $1-h((1+y)/2)\geq y^2/(2\ln{2})$ yields
\alg{
J_n:=
\sum_{i=0}^{N-1}[1-h(P_i)]
\geq 
\frac{\left[1+(\sqrt{2}-1)\tau\right]^{2n}}{2\ln{2}},
}
where  $h$ is the binary entropy defined by $h(x):=-x\log{x}-(1-x)\log{(1-x)}$.
Thus, we have $J_n>1$ for any $\tau>0$ and sufficiently large $n$.
On the other hand, as we have shown in the main text, the chain inequality of the mutual information \req{CI} implies $J_n\leq1$.
In the proof, we used the relation $I(X^N\!:\!MA^{N-1})\leq1$,
which follows from \req{italy} and \rLmm{omoida}.
Hence, we conclude that the chain inequality does not hold if $\tau>0$.

We construct the protocol achieving the success probability \req{supr} based on the {\it nested van Dam's protocol}, which was introduced in \cite{pawlowski09} to derive Tsirelson's bound from information causality.
In the protocol, the $(N-1)$ qubits are classified into $n$ layers in total.
We label the layers by $\alpha=0,\cdots,n-1$.
The $\alpha$-th layer is composed of an array $\vec{A}_\alpha$ of $2^\alpha$ qubits and an array $\vec{X}_{\alpha}$ of random bits of length $2^\alpha$.
We set $\vec{X}_{0}=M$ and $\vec{X}_n=\vec{X}$.
We denote by $X_{\alpha,\gamma}$ the $\gamma$-th bit in an array $\vec{X}_\alpha$, and by $A_{\alpha,\gamma}$ the $\gamma$-th component of $\vec{A}_\alpha$, where $\gamma=0,\cdots,2^\alpha-1$.
The bit array $\vec{X}_{\alpha+1}$ will be encoded to $\vec{A}_\alpha$ and will be encrypted by $\vec{X}_\alpha$.
In detail, for each $\alpha$ and $\gamma$, the pair of the bits $(x_{\alpha+1,2\gamma},x_{\alpha+1,2\gamma+1})$ is encoded in $A_{\alpha,\gamma}$ and is encrypted by $x_{\alpha,\gamma}$
by the map
\alg{
&
(x_{\alpha+1,2\gamma},x_{\alpha+1,2\gamma+1},x_{\alpha,\gamma})
\nn\\
&
\quad
\mapsto
\sigma_y^{x_{\alpha,\gamma}}\rho_{x_{\alpha+1,2\gamma},x_{\alpha+1,2\gamma+1}}\sigma_y^{x_{\alpha,\gamma}}
\nn\\
&\quad\quad\quad
=
\rho_{x_{\alpha+1,2\gamma}\oplus x_{\alpha,\gamma},x_{\alpha+1,2\gamma+1}\oplus x_{\alpha,\gamma}},
\!\!
\laeq{iiyatsu}
} 
where $\rho_{kl}$ is defined by \req{hashiru}.

To present the decoding scheme for $X_i$, let
\alg{
i=
\sum_{j=0}^{n-1}b_j\cdot2^{n-j-1}
}
be the binary decomposition of $k$, where $b_j\in\{0,1\}$.
Define also
\alg{
\gamma_\alpha=
\sum_{j=0}^{\alpha-1} b_j\cdot2^{\alpha-j-1}.
}
The decoder performs a measurement on $A_{\alpha,\gamma_\alpha}$ in each layer to make a guess at the bit $X_{\alpha+1,\gamma_{\alpha+1}}$.
The measurement is in terms of the basis $\{|k_z\rangle\}_{k=0,1}:=\{\ket{0},\ket{1}\}$ if $b_\alpha=0$ and $\{|l_x\rangle\}_{l=0,1}:=\{\ket{+},\ket{-}\}$ if $b_\alpha=1$.
Let $Y_{\alpha+1}$ be the outcome of the measurement on $A_{\alpha,\gamma_\alpha}$.
The probability of obtaining a measurement outcome $Y_{\alpha+1}$ depends only on $X_{\alpha+1,\gamma_{\alpha+1}}$ and $X_{\alpha,\gamma_\alpha}$, because of \req{iiyatsu} and
\alg{
\bra{k_z}\rho_{kl}\ket{k_z}
=
\bra{l_x}\rho_{kl}\ket{l_x}
=
1-\epsilon(\tau),
\laeq{kurea}
}
where $\epsilon(\tau):=(2-\sqrt{2})(1-\tau)/4$.
Thus, the probability is given by
\alg{
{\rm Pr}\{Y_{\alpha+1}=X_{\alpha+1,\gamma_{\alpha+1}}\oplus X_{\alpha,\gamma_\alpha}\}
=
\frac{1+\tau'}{2},
}
where
\alg{
\tau'=\frac{1+(\sqrt{2}-1)\tau}{\sqrt{2}}.
}
From $\alpha=0$ to $\alpha=n-1$ in order,
the decoder makes a guess $\hat{X}_{\alpha+1,\gamma_{\alpha+1}}$ at the bit $X_{\alpha+1,\gamma_{\alpha+1}}$ from $\hat{X}_{\alpha,\gamma_\alpha}$ and $Y_{\alpha+1}$ by
\alg{
\hat{X}_{\alpha+1,\gamma_{\alpha+1}}
=
\hat{X}_{\alpha,\gamma_\alpha}\oplus Y_{\alpha+1},
\quad
\hat{X}_{0,\gamma_0}=M.
}
The probability of correctly guessing $X_{\alpha+1,\gamma_{\alpha+1}}$ is then calculated by
\alg{
\kappa_{\alpha+1}
&:=
{\rm Pr}\{\hat{X}_{\alpha+1,\gamma_{\alpha+1}}=X_{\alpha+1,\gamma_{\alpha+1}}\}
\nn\\
&\quad\quad
-{\rm Pr}\{\hat{X}_{\alpha+1,\gamma_{\alpha+1}}\neq X_{\alpha+1,\gamma_{\alpha+1}}\}
\nn\\
&
=
\frac{1+\tau'}{2}
\cdot
{\rm Pr}\{\hat{X}_{\alpha,\gamma_{\alpha}}=X_{\alpha,\gamma_{\alpha}}\}
\nn\\
&\quad
+
\frac{1-\tau'}{2}\cdot
{\rm Pr}\{\hat{X}_{\alpha,\gamma_{\alpha}}\neq X_{\alpha,\gamma_{\alpha}}\}
\nn\\
&
\quad
-\frac{1+\tau'}{2}
\cdot
{\rm Pr}\{\hat{X}_{\alpha,\gamma_{\alpha}}\neq X_{\alpha,\gamma_{\alpha}}\}
\nn\\
&\quad
-
\frac{1-\tau'}{2}\cdot
{\rm Pr}\{\hat{X}_{\alpha,\gamma_{\alpha}}= X_{\alpha,\gamma_{\alpha}}\}
\\
&=\tau'\cdot\kappa_{\alpha}
\\
&=\tau'^{\alpha+1}\cdot\kappa_0.
}
Using $\kappa_0=1$ and $X_i=X_{n,\gamma_n}$, the probability of correctly guessing $X_i$ is obtained as
\alg{
P_i=\frac{1+\kappa_n}{2}=\frac{1+\tau'^n}{2},
}
which implies \req{supr}.

To prove that $I_{acc}(X^N\!:\!A^{N-1})=0$, we describe each bit $X_{\alpha,\gamma}$ by $\mbb{C}^2$.
Since it is a classical system, we may assume that the states are diagonal with respect to a fixed orthonormal basis $\{\ket{x}\}_{x=0,1}$.
Observe that the encoding map \req{iiyatsu} is represented by a completely positive trace-preserving map
$\ca{E}_{\alpha,\gamma}$ from $X_{\alpha+1,2\gamma}X_{\alpha+1,2\gamma+1}$ to $A_{\alpha,\gamma}X_{\alpha,\gamma}$ defined by
\alg{
&
\proj{x_{\alpha+1,2\gamma}}\otm\proj{x_{\alpha+1,2\gamma+1}}
\nn\\
&
\quad\quad\quad
\mapsto
\frac{1}{2}
\!\sum_{x_{\alpha,\gamma}=0,1}
\!\!\!
\sigma_y^{x_{\alpha,\gamma}}\rho_{x_{\alpha+1,2\gamma},x_{\alpha+1,2\gamma+1}}\sigma_y^{x_{\alpha,\gamma}}
\nn\\
&\quad\quad\quad\quad\quad\quad\quad\quad\quad\quad\quad\quad
\otm
\proj{x_{\alpha,\gamma}}.
}
It follows from \req{iiyatsu} and \req{kyos} that
\alg{
{\rm Tr}_{X_{\alpha,\gamma}}
\circ
\ca{E}_{\alpha,\gamma}
=
\pi^{A_{\alpha,\gamma}}
\cdot
{\rm Tr}_{X_{\alpha+1,2\gamma}X_{\alpha+1,2\gamma+1}},
\laeq{aiteni}
}
where $\pi$ is the maximally mixed state.
The encoding operation in the $\alpha$-th layer is simply given by
$
\tilde{\ca{E}}_{\alpha}
=
\ca{E}_{\alpha,0}\otm\cdots\otm\ca{E}_{\alpha,2^\alpha-1}
$,
and the whole encoding map is
$
\tilde{\ca{E}}
=
\tilde{\ca{E}}_{0}\circ\cdots\circ\tilde{\ca{E}}_{n-1}
$.
Abbreviating $|x^N\rangle\!\langle x^N|$ by $x^N$,
the state of the system $MA^{N-1}$ after encoding the bit array $x^N$ is given by
\alg{
\rho_{x^N}^{MA^{N-1}}
:=
\tilde{\ca{E}}(x^N).
}
Due to the relation \req{aiteni}, it holds that
\alg{
\rho_{x^N}^{A^{N-1}}
&
=
{\rm Tr}_M\circ\tilde{\ca{E}}(x^N)
\\
&
=
{\rm Tr}_M\circ\tilde{\ca{E}}_{0}\circ\cdots\circ\tilde{\ca{E}}_{n-1}(x^N)
\\
&=\pi^{A_{0,0}}\cdot {\rm Tr}_{\vec{X}_0}\circ\tilde{\ca{E}}_{1}\circ\cdots\circ\tilde{\ca{E}}_{n-1}(x^N)
\\
&=\pi^{\vec{A}_{0}}\otm\pi^{\vec{A}_{1}} \cdot {\rm Tr}_{\vec{X}_1}\circ\tilde{\ca{E}}_{2}\circ\cdots\circ\tilde{\ca{E}}_{n-1}(x^N)
\\
&=\cdots
\\
&=\pi^{A^{N-1}} \cdot {\rm Tr}_{\vec{X}}(x^N)
\\
&=\pi^{A^{N-1}}.
}
Thus, the state $\rho_{x^N}^{A^{N-1}}$ does not depend on the value of $x^n$.
This implies $I_{acc}(X^N\!:\!A^{N-1})=0$ and completes the proof.
\QED

\section{Proofs of Lemmas}
\lapp{prfs}

{\bf Proof of \rLmm{CGmeasdec}:}
To prove \req{iwasok}, recall the assumptions that there exists an operation $\ca{E}_x\in\mathfrak{O}(A\rightarrow XA)$ satisfying $
\ca{E}_x(\phi)=e_x\times \phi$.
Let $\nu_x\in\mathfrak{M}(A)$ be a measurement composed of the operation $\ca{E}_x$ and the subsequent measurement $\nu\in\mfk{M}(XA)$.
By definition, we have $\mathfrak{R}(\nu_x)=\mathfrak{R}(\nu)$ and for any $\phi\in\mathfrak{S}(A)$, it holds that
\alg{
p(s|e_x\times\phi,\nu)=
p(s|\phi,\nu_x).
}
From \req{rhodfn}, we have
\alg{
p(s|\rho,\nu)
&
=
\sum_{x\in\ca{X}}p(x)p(s|e_x\times\phi_x,\nu)
\nn\\
&
=
\sum_{x\in\ca{X}}p(x)p(s|\phi_x,\nu_x),
\laeq{iwaiwa}
}
which implies \req{iwasok}.
Let $\mu_X$ be a measurement on $X$ such that $p(x'|e_x,\mu_X)=\delta_{x,x'}$.
From the assumption of the existence of sequential measurements, there exists a measurement $\tilde{\nu}\in\mathfrak{M}(XA)$ such that for any $\rho\in\mathfrak{S}(XA)$, it holds that
\alg{
p(x,s|\rho,\mu_X\times\nu_x)
=
p(x,s|\rho,\tilde{\nu}).
}
The L.H.S. is calculated for $\rho= e_{x'}\times\phi$ to be
\alg{
p(x,s|e_{x'}\times\phi,\mu_X\times\nu_x)
=
\delta_{x,x'}
\cdot
p(s|\phi,\nu_x),
}
which implies \req{iwasok3}.
It follows that
\alg{
&
\sum_x
p(x,s|\rho,\mu_X\times\nu_x)
\nn\\
&
=
\sum_x\sum_{x'}
p(x')
\delta_{x,x'}
\cdot
p(s|\phi_x,\nu_x)
\nn\\
&
=
\sum_x
p(x)
\cdot
p(s|\phi_x,\nu_x).
}
Combining this with \req{iwaiwa}, we obtain \req{iwasok2}.
\QED

\hfill

{\bf Proof of \rLmm{TDCG}:}
Recall that the Kolmogorov distance is defined by
\alg{
d(\phi,\psi)
:=
\sup_{\mu\in\mathfrak{M}(A)}\frac{1}{2}\sum_{r\in\mathfrak{R}(\mu)}\!\left|p(r|\phi,\mu)-p(r|\psi,\mu)\right|.
}
Due to the assumption that the set of measurements is closed under classical post processing, without loss of generality we may assume that $\mu$ is a binary measurement i.e., $\mathfrak{R}(\mu)=\{0,1\}$.
In this case, we may further assume that $p(r=0|\phi,\mu)-p(r=0|\psi,\mu)\geq0$.
This leads to
\alg{
\!
d(\phi,\psi)
=
\sup_{\mu\in\mathfrak{M}_{\rm bi}(A)}[p(r=0|\phi,\mu)-p(r=0|\psi,\mu)],
\!
\laeq{tracedistancebinary}
}
where the supremum is taken over all binary measurements on $A$.

To prove \req{dRS1}, we first show that
\alg{
d(\rho,\sigma)
\geq
\sum_{x\in\ca{X}}p_xd(\phi_x,\psi_x).
\laeq{TDcqdec}
}
For each $x$, let $\nu_x$ be the binary measurement such that 
\alg{
d(\phi_x,\psi_x)
=
[p(s=0|\phi_x,\nu_x)-p(s=0|\psi_x,\nu_x)].
}
Due to \rLmm{CGmeasdec}, there exists a binary measurement $\nu\in\mathfrak{M}(XA)$ such that
\alg{
&
p(s|\rho,\nu)
=
\sum_{x\in\ca{X}}p(x)p(s|\phi_x,\nu_x),
\laeq{ink1}
\\
&
p(s|\sigma,\nu)
=
\sum_{x\in\ca{X}}p(x)p(s|\psi_x,\nu_x).
\laeq{ink2}
}
This implies
\alg{
&
\!\!\!\!\!\!
d(\rho,\sigma)
\geq
p(s=0|\rho,\nu)
-
p(s=0|\sigma,\nu)
\\
&
=
\sum_{x\in\ca{X}}p(x)
[p(s=0|\phi_x,\nu_x)
-
p(s=0|\psi_x,\nu_x)]
\\
&
=
\sum_{x\in\ca{X}}p(x)
d(\phi_x,\psi_x),
}
which leads to \req{TDcqdec}.
Second, we prove that
\alg{
d(\rho,\sigma)
\leq
\sum_{x\in\ca{X}}p_xd(\phi_x,\psi_x).
\laeq{TDcqdec2}
}
Let $\nu$ be the binary measurement such that
\alg{
d(\rho,\sigma)
=
[p(s=0|\rho,\nu)-p(s=0|\rho,\nu)].
}
Due to \rLmm{CGmeasdec}, there exists a set of measurements $\{\nu_x\}$ on $XA$ such that Inequalities \req{ink1} and \req{ink2} hold.
We have
\alg{
&
\!\!\!\!\!\!
d(\rho,\sigma)
=
p(s=0|\rho,\nu)
-
p(s=0|\sigma,\nu)
\\
&
=
\sum_{x\in\ca{X}}p(x)
[p(s=0|\phi_x,\nu_x)
-
p(s=0|\psi_x,\nu_x)]
\\
&
\leq
\sum_{x\in\ca{X}}p(x)
d(\phi_x,\psi_x).
}
This implies \req{TDcqdec2} and completes the proof.
\QED

\hfill\\
\noindent
{\bf Proof of \rLmm{omoida}:}
Inequality \req{hirokumo} is proved in \cite{wakakuwa2012chain} (See Lemma 6.1 therein).
To prove \req{migaki}, fix arbitrary $R>H(X|TA)_\rho$, $\epsilon>0$ and choose sufficiently large $n$.
By definition, there exists a function $f:\ca{X}^n\rightarrow[2^{nR}]$ and a set of measurements $\{\mu_{x',t^n}\}_{x'\in[2^{nR}],t^n\in\ca{T}^n}$ on $A^n$ such that
\alg{
\sum_{x^n\in \ca{X}^n,t^n\in \ca{T}^n}p(x^n|\phi_{x^n,t^n}^n,\mu_{x'=f(x^n),t^n})p(x^n,t^n)\geq1-\epsilon.
}
Note here that any measurement on $T^nA^n$ is represented as a sequential measurement from $T^n$ to $A^n$ (\rLmm{CGmeasdec}).
Let $M\equiv f(X^n)$ and let $\hat{X}^n$ be a random variable that represents the result of the measurement $\mu_{x',t^n}$.
Due to Fano's inequality \cite{cover05}, it holds that
\alg{
I(X^n:MT^n\hat{X}^n)
\geq
nH(X)-n\epsilon\log{|\ca{X}|}-h(\epsilon),
}
where $h$ is the binary entropy defined by $h(x):=-x\log{x}-(1-x)\log{(1-x)}$.
The chain rule of the classical mutual information implies
\alg{
I(X^n\!:\!MT^n\hat{X}^n)
=
I(X^n\!:\!M)
\!+\!
I(X^n\!:\!T^n\hat{X}^n|M).
\!
}
Noting that $M=f(X^n)$, the first term is bounded as
\alg{
I(X^n:M)
\leq
H(M)
\leq
nR.
}
In addition, we have
\alg{
I(X^n:T^n\hat{X}^n|M)
\leq
H(T^n)
}
from Inequality \req{hirokumo}.
This is because the state $\rho^{\times n}$ is decoupled between $X^n$ and $A^n$ for each value of $M$ unless $T^n$ is given, due to \req{anohito}.
Combining these all together, we obtain
\alg{
R+H(T)
\geq
H(X)
-
\epsilon\log{|\ca{X}|}-\frac{h(\epsilon)}{n}.
}
Since this relation holds for any $R>H(X|TA)_\rho$ and $\epsilon>0$, we arrive at \req{migaki}.
\QED

\section{A Refinement of The Uncertainties Relation}
\lapp{extUR}

In the main text, we have formulated the uncertainties relation and proved it under the assumption of GMP.
However, it does not properly capture the relation between uncertainties if the pair of observables are almost or exactly compatible.
In the following, we provide a refinement of the uncertainties relation to overcome this limitation.

Consider a pair of measurements $\mu,\nu\in\mfk{M}(A)$.
Suppose that the sets of the outcomes of the measurements are decomposed into the same number of disjoint subsets as
\alg{
\mfk{R}(\mu)
&
=
\mfk{R}_1(\mu)\cup\cdots\cup\mfk{R}_\Omega(\mu),
\laeq{decRN1}
\\
\mfk{R}(\nu)
&
=
\mfk{R}_1(\nu)\cup\cdots\cup\mfk{R}_\Omega(\nu),
\laeq{decRN2}
}
where $\Omega\in\mbb{N}$.
Suppose also that for any $\omega\in[\Omega]$ and any pair of measurement results $x\in\mfk{R}_\omega(\mu)$ and $y\in\mfk{R}_\omega(\nu)$, there exists a state $\phi_{xy}\in\mfk{S}(A)$ and it holds that
\alg{
p(x|\phi_{xy},\mu)
\geq
1-\epsilon,
\quad
p(y|\phi_{xy},\nu)
\geq
1-\epsilon,
\laeq{crycry}
}
where $\epsilon\in(0,1]$.
From the assumption of GMP, we can show that there exists a measurement $\xi\in\mfk{M}(A)$ such that $\mfk{R}(\xi)\!=\!\mfk{R}(\mu)\!\times\!\mfk{R}(\nu)$ and it holds that
\alg{
p((x,y)|\phi_{xy},\xi)
\geq
1-2\eta(\epsilon)
}
for any $\omega\in[\Omega]$ and $(x,y)\in\mfk{R}_\omega(\mu)\!\times\!\mfk{R}_\omega(\nu)$.
The proof is along the same line as in the case of $\Omega=1$ in the main text.
The uncertainties relation in this form incorporates the case where the observables are almost or exactly compatible, 
as we may choose the decomposition \req{decRN1} and \req{decRN2} depending on the observables.

\end{document}